\documentclass[aps,showpacs]{revtex4}
\usepackage{graphicx,psfrag}
\def\bc{\begin{center}}
\def\ec{\end{center}}

\def\beq{\begin{equation}}
\def\eeq{\end{equation}}

\begin{document}

\title{
Controlling dynamical entanglement in a Josephson tunneling junction
}

\author{K. Ziegler}
\affiliation{Institut f\"ur Physik, Universit\"at Augsburg, D-86135 Augsburg, Germany
}

\begin{abstract}
We analyze the evolution of an entangled many-body state in a Josephson
tunneling junction. A N00N state, which is a superposition of two complementary 
Fock states, appears in the evolution with sufficient probability only for a moderate
many-body interaction on an intermediate time scale. This time scale is inversely proportional 
to the tunneling rate. Interaction between particles supports entanglement: The probability for creating an entangled 
state decays exponentially with the number of non-interacting particles, whereas it decays only 
like the inverse square root of the number of interacting particles.
\end{abstract}
\pacs{03.65.Aa, 03.65.Fd, 03.67.Bg}

\maketitle

\section{Introduction}

In quantum optics, multiphoton entangled states can be used to carry out high-precision measurements. 
A good candidate is the N00N state \cite{sanders89}
which has attracted much attention for highly accurate interferometry and
other precision measurements \cite{lee02,walther04,mitchell04,afek10,israel11}.
The method becomes even more advantageous as the number of photons grows. 
There exist several protocols for the dynamical creation of entangled states. An early suggestion was
to exploit the coupling of photons to a transparent medium with a strong Kerr 
nonlinearity \cite{yurke86}. This idea was followed by a number of other proposals, based on beam splitters
and anharmonic photon effects \cite{gerry01,paternostro04,nielsen07,pezze08,platzer10,liu08}.
The number of photons is limited, though, especially by the harmonic effect of beam splitters, 
such that the probability for creating a N00N state decreases exponentially with the number of photons $N$ \cite{kok02}. 

Alternative systems for the creation of N00N states are nuclear spins \cite{jones09}
and atomic systems \cite{chen10}. Of particular interest is a gas of ultracold bosonic atoms
in a double well, where the potential wells are created by a Laser field \cite{weiss10}. 
This would be a realization of a Josephson
tunneling junction, where entangled atomic states can be created \cite{ketterle04,oberthaler05}. 
A double-well potential, filled with ultracold atoms, has become a standard experimental
set up in which many-body quantum states can be studied in great detail. 
The experimental conditions are such that the atoms can be considered
as isolated from the environment for the time of the experiment. Therefore,
besides photons in optical cavities \cite{cqed,haroche}, trapped ultracold bosonic atoms provide
a playground to analyze the complex evolution of many-body states, including
squeezing, entanglement and correlation effects \cite{atoms}. 

\begin{figure}
\psfrag{H}{$H$}
\psfrag{H_0}{$H_0$}
\psfrag{H_1}{$H_1$}
\begin{center}
\includegraphics[width=9cm,height=5cm]{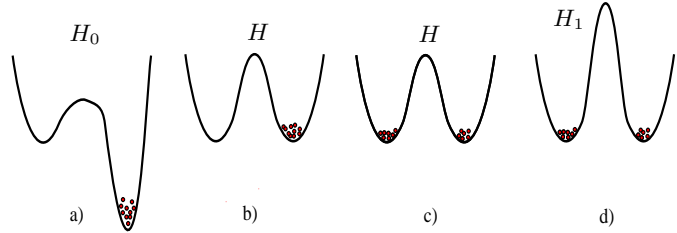}
\caption{
Procedure for the preparation of an entangled state in a double well: a) preparation of a Fock state $|0,N\rangle$
in the right well. a) $\to$ b): A sudden change from the asymmetric to a symmetric double well. 
This implies a change of the Hamiltonian $H_0\to H$. b) $\to$ c): The evolution of the quantum state 
under the influence of the Hamiltonian $H$. c) $\to$ d): A sudden change of the barrier potential between
the two wells suppresses tunneling. It results in a change of the Hamiltonians $H\to H_1$, where the entanglement
is preserved.
}
\label{fig:double_well}
\end{center}
\end{figure}

For a closed quantum system, the unitary evolution $\exp(-iHt)|\Psi_0\rangle$
with a Hamiltonian $H$ provides a N00N state at a given time $t$ when the
initial state is $|\Psi_0\rangle=\exp(iHt)|N00N\rangle$.
($H$ is a Hamiltonian normalized by $\hbar$, which implies that it has the dimension of 
a frequency.) In a real situation, though, it is not possible to generate an arbitrary initial state.
A natural candidate for the initial state of an atomic system in a trapping potential is a Fock state.
Guided by recent experiments with ultracold atoms we propose a procedure for the
dynamical creation of entangled states. For this purpose we consider a double well potential 
whose potential shape is manipulated by external Laser fields.
There are three major aspects, which must be distinguished in the experiment:
First is the preparation of a well-defined initial state, then there is its evolution
for a period of time due to quantum tunneling and particle-particle interaction. 
And finally, the evolution must be stopped at a certain time $t_e$ and the resulting state must
be kept in the double well without tunneling between the wells. For the 
preparation of the initial state one can use the ground state of a special Hamiltonian $H_0$.
This could be an asymmetric double-well potential as illustrated in Fig. \ref{fig:double_well} a)
for the preparation of a Fock state $|0,N\rangle\equiv |0\rangle\otimes|N\rangle$.
Then the evolution with $\exp(-iHt)$ 
for a different Hamiltonian $H$ involves a sudden change $H_0\to H$
with a simultaneous reduction of the barrier between the wells \cite{ketterle04,oberthaler05,trotzky08}. 

The sudden change of the Hamiltonian is usually called a quench. An example is shown in Fig. \ref{fig:double_well} b), 
where the potential becomes symmetric. The asymmetric Fock state
$|0,N\rangle$ is not an eigenstate of the new Hamiltonian $H$, which causes an evolution
$|\Psi_t\rangle=  \exp(-iHt)|0,N\rangle$ inside the Hilbert space that is spanned by the
states $\{|k,N-k\rangle\}_{0\le k\le N}$ (cf. Fig. \ref{fig:double_well} c)).
The evolution from the initial Fock state can, in principle, lead to an entangled state, such as the
state $c_0|0,N\rangle +c_N|N,0\rangle$. This entangled state resembles the N00N state 
$|N00N\rangle=\left(|N,0\rangle +e^{i\phi}|0,N\rangle\right)/\sqrt{2}$. The degree of entanglement
changes over time. Therefore, we must stop the evolution at a proper time $t_e$ by stopping the tunneling
between the potential wells.
This can be achieved by simply raising the potential barrier between the wells, as indicated in
Fig. \ref{fig:double_well} d). The evolution must be analyzed in detail to determine an optimal 
entanglement time $t_e$. An important aspect is the scaling behavior with the number of particles $N$.
To avoid problems with uncontrolled approximations, we will rely on a 
full quantum calculation. An exact solution is available from a directed walk through the Hilbert space 
in a Fock-state base, as described previously in Ref. \cite{ziegler11}.


The paper is organized as follows: In Sect. \ref{sect:model} we define the Bose-Hubbard model as a description
of a Josephson tunneling junction for interacting bosons and several quantities which provide a measure
for the dynamical creation of nearly N00N states. Next, in Sect. \ref{sect:evolution} the evolution of the
pure Fock state $|0,N\rangle$ is discussed, where we start with the simple case of non-interaction bosons
(Sect. \ref{sect:non_int}) and then consider interacting bosons (Sect. \ref{sect:inter}). Finally, we
suppress tunneling between the potential wells and study the evolution of the resulting entangled state
in Sect. \ref{sect:no_tunn}.


\section{Model}
\label{sect:model}

A Josephson tunneling junction, represented by a bosonic system in a double well, is described in two-mode approximation by the 
Bose-Hubbard Hamiltonian \cite{oberthaler07}
\beq
H=-\frac{J}{2}(a_l^\dagger a_r + a_r^\dagger a_l)+ \frac{U}{2}(n_l^2+n_r^2) , \ \ \ 
n_{l,r}=a_{l,r}^\dagger a_{l,r}
\ ,
\label{ham00}
\eeq
where $a^\dagger_{l,r}$ ($a_{l,r}$) are the creation (annihilation) operators in the left and right potential well, 
respectively. 
The first term of $H$ describes tunneling of atoms between the wells, and for $U>0$ the second 
term represents a repulsive particle-particle interaction that favors energetically
a symmetric distribution of bosons in the double well. Without tunneling (i.e., for $J=0$)
the eigenstates with energy $E_k=U[(N-k)^2+k^2]/2$ are superpositions of $|k,N-k\rangle$ and $|N-k,k\rangle$. 

In the following we will focus on a unitary evolution of the initial state, where all bosonic atoms are 
located in the right potential well as the product Fock state $|0,N\rangle$: 
$|\Psi_t\rangle=e^{-iHt}|0,N\rangle$.
Then we determine the return amplitude to the initial state $c_0=\langle 0,N|\Psi_t\rangle$ and the transition 
amplitude $c_N=\langle N,0|\Psi_t\rangle$. With these two amplitudes it is possible to calculate 
the overlap of $|\Psi_t\rangle$ with the N00N state
$|N00N\rangle=\left(|N,0\rangle +e^{i\phi N}|0,N\rangle\right)/\sqrt{2}$ as
\beq
\langle N00N|\Psi_t\rangle
=\frac{c_0+e^{-i\phi N} c_N}{\sqrt{2}}
\ .
\eeq
The phase $\phi$ of the N00N state can be measured as the expectation value of the operator 
$A=|N,0\rangle\langle 0,N|+|0,N\rangle\langle N,0|$ as
\cite{kok02}
\beq
\langle N00N|A|N00N\rangle=\cos \phi
\ .
\label{phase0}
\eeq
The evolution of the Fock state $|0,N\rangle$ will not create a pure N00N state but a state
\beq
|\Psi_t\rangle=c_0|0,N\rangle + c_N|N,0\rangle +\sum_{k=1}^{N-1}c_k|k,N-k\rangle \ , \ \ \ 
A|\Psi_t\rangle=c_N|0,N\rangle + c_0|N,0\rangle
\ ,
\eeq
which represents a pseudo N00N state (PNS) if $c_0,c_N\ne 0$ simultaneously. Then, instead of (\ref{phase0}), we
get a phase-sensitive expression of the PNS as
\beq
\langle\Psi_t|A|\Psi_t\rangle=2Re(c_0^* c_N)
\ .
\label{coherence0}
\eeq
For a N00N state with $\phi=0$ this expectation value is the coherent part of the Husimi--$Q$ function \cite{haroche}:
\beq
Q=\frac{1}{\pi}|\langle N00N|\Psi_t\rangle|^2
=\frac{1}{2\pi}\left[|c_0|^2+|c_N|^2+2Re( 
c_0^* c_N)\right]
\ ,
\label{husimi0}
\eeq
which is proportional to the fidelity $|\langle N00N|\Psi_t\rangle|^2$.
Moreover, $A^2$ is the projector onto the space spanned by $\{|0,N\rangle, |N,0\rangle\}$: 
\[
A^2|\Psi_t\rangle=c_0|0,N\rangle + c_N|N,0\rangle
\ ,
\]
such that $\langle\Psi_t|A^2|\Psi_t\rangle=|c_0|^2+|c_N|^2$ 
\beq
Q=\frac{1}{2\pi}\left(\langle\Psi_t|A^2|\Psi_t\rangle + \langle\Psi_t|A|\Psi_t\rangle \right)
\ .
\eeq
In the following we will study the behavior of the coefficients $c_0$ and $c_N$ as functions of time $t$ and the
resulting expectation value $\langle\Psi_t|A|\Psi_t\rangle$. Moreover, to distinguish the PNS from the pure Fock states, 
we introduce the probability for the creation of a PNS as $p_e(t)=2|c_0c_N|$ and its maximum with respect to time
\beq
P_e=2\max_t|c_0c_N|
\ .
\label{ent00}
\eeq

\section{Evolution}
\label{sect:evolution}

We consider a unitary evolution with the initial state $|\Psi_0\rangle$  
and obtain with the Hamiltonian $H$ for the return amplitude $c_0$ and the transition amplitude $c_j$ as
\beq
c_0=\langle\Psi_0|\Psi_t\rangle=\langle\Psi_0|e^{-iHt}|\Psi_0\rangle , \ \ \
c_j=\langle\Psi_j|\Psi_t\rangle=\langle\Psi_j|e^{-iHt}|\Psi_0\rangle
\ .
\label{evol_coeff}
\eeq
Starting from these amplitudes, a Laplace transformation relates them with the resolvent through the identity
\beq
\langle\Psi_j|\Psi_t\rangle 
=\int_\Gamma \langle\Psi_j|(z-H)^{-1}|\Psi_0\rangle e^{-izt}dz \ \ \ (j=0,1)
\ ,
\eeq
where the contour $\Gamma$ encloses all the eigenvalues $E_k$ ($k=0,1,...,N$) of $H$, assuming that
the underlying Hilbert space is $N+1$ dimensional.
With  the corresponding eigenstates $|E_k\rangle$ the spectral representation of the resolvent is 
a rational function of $z$:
\beq
\langle\Psi_j|(z-H)^{-1}|\Psi_0\rangle=\sum_{k=0}^N\frac{|\langle\Psi_j|E_k\rangle\langle E_k|\Psi_0\rangle}{z-E_k}
=\frac{P_{j,N}(z)}{Q_{N+1}(z)}, \ \ \
Q_{N+1}(z)=\prod_{k=0}^N (z-E_k)
\ ,
\label{resolvent2}
\eeq
where $P_{j,N}(z)$, $Q_{N+1}(z)$ are polynomials in $z$ of order $N$, $N+1$, respectively. 
These polynomials can be evaluated by the 
recursive projection method (RPM) \cite{ziegler11}. This method is based on a systematic expansion of
the resolvents $\langle \Psi_j|(z-H)^{-1}|\Psi_0\rangle$, starting from the initial state $|\Psi_0\rangle$.
It can be understood as a directed walk in Hilbert space, where each subspace
is only visited once. 
This is the main advantage of the RPM that allows us to calculate efficiently the resolvent 
$\langle \Psi_j|(z-H)^{-1}|\Psi_0\rangle$ on an $N+1$-dimensional Hilbert space. Results are discussed and
visualized in the next section for non-interacting as well as interacting bosons in a Josephson tunneling junction.

\subsection{Non-interacting bosons}
\label{sect:non_int}

We begin our discussion with non-interacting bosons (i.e., for $U=0$), where the Hamiltonian describes unrestricted tunneling
between the wells. Such a system can be realized for photons at a beam splitter \cite{yurke86,haroche} or in two harmonic cavities, 
which are connected through
an optical fiber \cite{ziegler12}. For $N$ particles ($N$ even), the $N+1$ energy levels 
$E_k=-J(N/2-k)$ ($k=0,1,...,N$) are equidistant with eigenstates 
\[
|E_k\rangle=\frac{2^{-N/2}}{\sqrt{k!(N-k)!}}(a_l^\dagger+a_r^\dagger)^k(a_l^\dagger-a_r^\dagger)^{N-k}|0,0\rangle
\ .
\]
Thus, the fastest oscillations occur with frequency $NJ/2$. The behavior of $c_0$ and $c_N$ is much smoother, 
though, with frequency $J/2$:
\beq
c_0 
=\cos^N(Jt/2) , \ \ \
c_N 
=(-i)^N\sin^N(Jt/2)
\ .
\label{non_int_exp}
\eeq
For short times (i.e., for $J t\ll 1/\sqrt{N}$) we have a Gaussian decay of the Fock state
\beq
|c_0|=|\cos^N(Jt/2)|\sim e^{-J^2Nt^2/8}
\ .
\label{decay0}
\eeq
This decay, which differs from the exponential decay of classical systems, is related to the quantum Zeno effect
\cite{misra77}.
The measure for entanglement $p_e(t)=2|c_0c_N|$ reads for even $N$
\beq
p_e(t) 
=\frac{\sin^N(Jt)}{2^{N-1}}
\ ,
\label{ent1}
\eeq
such that the maximum with respect to time in (\ref{ent00}) is
\beq
P_e=2^{-(N-1)}
\ ,
\label{ent0}
\eeq
describing an exponential decay with the number of particles. 
This is also the case for the coherent part of the Husimi--$Q$ function (\ref{husimi0})
\[
Q=\frac{1}{2\pi}\left[\cos^{2N}(Jt/2)+\sin^{2N}(Jt/2)+\cos(N\pi/2)\frac{\sin^N(Jt)}{2^{N-1}}\right]
\]
and for the expectation value of $A$
\beq
\langle\Psi_t|A|\Psi_t\rangle = 
\cos(N\pi/2)\frac{\sin^N(Jt)}{2^{N-1}}
\ .
\label{A_non}
\eeq
The protocol, suggested for the creation of 
N00N states consisting of photons in Ref. \cite{kok02}, also has an exponential decay with $N$.
This implies that the creation of a PNS is only possible for a small number of particles. 

The periodic dynamics of $|c_0|$ and $|c_N|$, as defined in (\ref{non_int_exp}), is plotted for $N=100$ 
particles  on the left-hand side of Fig. \ref{fig:dynamics1}. Without interaction 
the evolution of the initial 
Fock state $|0,N\rangle$ is periodic with period $2\pi/J$, according to Eq. (\ref{non_int_exp}). For not too small $N$ 
the state $|0,N\rangle$ decays quickly, according to Eq. (\ref{decay0}), and after the time period $\pi/J$
a Fock state appears in the other well. This state disappears quickly again and the system returns after another period $\pi/J$ to the
initial Fock state, as visualized on the left-hand side of Fig. \ref{fig:dynamics1}. Thus, there is an anti-correlation effect: 
$c_0$ vanishes when $c_N$ becomes nonzero and vice versa. This effect increases with $N$, reflecting the fact that in the classical
limit $N\to\infty$ there is no entanglement. The probability for the creation of the PNS 
is related to $p_e(t)$, which decays exponentially with $N$, as given in Eq. (\ref{ent1}). 
Therefore, the maximal value $P_e$ is strongly suppressed. 
This is a consequence of the fact that for an increasing $N$ the particles disappear in the $(N+1)$--dimensional Hilbert
space without contributing to the PNS because there is no constraint enforcing interaction.

\subsection{Interacting bosons}
\label{sect:inter}


The evolution with interaction $U>0$ experiences scattering between particles, which forces them 
into subspaces. This effect is related to Hilbert-space localization \cite{cohen15}, 
which can be understood from two asymptotic regimes: Our results for non-interacting
particles in Sect. \ref{sect:non_int} represent a periodic dynamics, in which all particles move freely to all
states. On the other hand, for strongly interacting particles the product
Fock states are almost eigenstates. This implies that the evolution will stay close to the initial
state $|0,N\rangle$ and may return quickly to it. This is indeed what we see in Figs. \ref{fig:dynamics1} 
and \ref{fig:dynamics2}, where we compare a non-interacting with interacting systems of increasing interaction
strength $u=NU/J$.
On the left-hand side of Fig. \ref{fig:dynamics1} we see the dynamics of $|c_0|$ and $|c_N|$ for $N=100$ 
particles with $u=0$, and on the right-hand side the dynamics with interaction $u=3.75$. The plots clearly 
indicate a substantial difference between the two cases: While non-interacting particles occupy the
states $|0,N\rangle$ and $|N,0\rangle$ at different times, this is not so for interacting particles.
The latter begin to occupy both states $|0,N\rangle$ and $|N,0\rangle$ at the
same time when the non-interacting particles begin to occupy the state $|N,0\rangle$ only. 
In other words, the return to the initial state $|0,N\rangle$ happens much earlier for the 
interacting particles, while the system still partially occupies the state $|N,0\rangle$.
This behavior is even more pronounced when we increase the interaction to $u=4$, which is
depicted on the left-hand side of Fig. \ref{fig:dynamics2}. In this case the evolution of the
two states $|0,N\rangle$ and $|N,0\rangle$ is almost simultaneous from $t=0.8$ sec to $t=1.6$ sec.
(For general tunneling rate this time interval is from $t=\pi/J$ to $t=2\pi/J$.)
A further increase of the interaction strength to $u=4.25$ indicates that the system prefers
to stay closer to the initial state rather than moving to $|N,0\rangle$ (right-hand side of Fig. 
\ref{fig:dynamics2}). Comparing these three cases we conclude that there is an optimum for the 
simultaneous occupation of both states $|0,N\rangle$ and $|N,0\rangle$ at $u=4$. 

The qualitative description is also supported by the behavior of the measure $p_e(t)$ for the creation of the PNS.
For the optimal interaction strength $u=4$ it is plotted for different particle numbers $N$ on the left-hand side
of Fig. \ref{fig:dynamics3}: $p_e(t)$ decreases with an increasing number of particles, at least for $t<1.3$ sec. 
The decrease, 
though, is not exponential as in the non-interacting case (cf. Eq. ({\ref{ent0})) but follows a power law. 
The maximal value $P_e$ is plotted as a function of $N$ on the right-hand side of Fig. \ref{fig:dynamics3}
and indicates a decay according to
\beq
P_e\sim 1.7 N^{-1/2}
\ .
\eeq
The expectation values $\langle\Psi_t|A|\Psi_t\rangle$ is not small even for $N=100$ particles if $u=4$
(cf. Fig. \ref{fig:dynamics4}), in comparison to the case $u=0$ of Eq. (\ref{A_non}).


\subsection{Blocked tunneling}
\label{sect:no_tunn}

A sudden increase of the barrier between the wells, as indicated in step c) $\to$ d) of Fig. \ref{fig:double_well}, 
results in a change of the Hamiltonian
\beq
H\to H_1=\frac{U_l}{2}n_l^2+\frac{U_r}{2}n_r^2
\ ,
\eeq
where $H_1$ has no tunneling term. Here we have included a simultaneous change of
the interaction parameter to different values in the wells. 
Such a change can be realized either by modifying the trapping potential \cite{ketterle04,oberthaler05,oberthaler07}
or by coupling photons directly to the atoms in the junction \cite{rosson15}.
The evolution after the raise of the barrier reads
\beq
|\Psi_t\rangle=e^{-iH_1t}\sum_{k=0}^N c_k|k,N-k\rangle
=\sum_{k=0}^N c_k e^{-iU_l k^2 t/2}|k\rangle\otimes e^{-iU_r (N-k)^2 t/2}|N-k\rangle
\ ,
\label{final_state}
\eeq
where the coefficients $c_k$ depend on the fixed time $t_e$ when the barrier was raised.
We can also consider the coefficients with time-dependent phase factors for times $t>t_e$:
\beq
c_0'=c_0e^{-iU_r N^2 t/2} , \ \ \ c_N'=c_Ne^{-iU_l N^2 t/2}
\ ,
\label{eff_coeff}
\eeq
which give a constant value for $p_e=2|c_0c_N|$, since $c_0$ and $c_N$ have been fixed at $t=t_e$.
The expectation value of $A$ is time dependent:
\beq
\langle\Psi_t|A|\Psi_t\rangle = 2Re\left(c_0^*c_Ne^{-i(U_l-U_r) N^2 t/2}\right)
=2|c_0c_N|\cos[-\varphi_0+(U_l-U_r) N^2 t/2]
\ ,
\label{expect_2}
\eeq
where $\varphi_0$ is the relative phase between $c_0$ and $c_N$. It describes a periodic behavior
with periodicity $T=4\pi/(U_l-U_r) N^2$.
The Husimi--$Q$ function with respect to the N00N state is also time dependent with
\beq
\frac{1}{\pi}|\langle N00N|\Psi_t\rangle|^2=\frac{1}{2\pi}|c_0'+e^{i\phi N}c_N'|^2
=\frac{1}{2\pi}|c_0+e^{i\phi N}e^{-i(U_l-U_r) N^2 t/2}c_N|^2
\ .
\eeq
By fixing $|c_0|$ and $|c_N|$ as $|c_0|=|c_N|$ in the tunneling dynamics
with the optimal parameter $u=4$ and proper time period $t_e$ (cf. left-hand side of Fig. \ref{fig:dynamics2})
the Husimi--$Q$ function becomes
\beq
\frac{1}{\pi}|\langle N00N|\Psi_t\rangle|^2
=\frac{|c_0|^2}{\pi}\left\{1+\cos\left[(N\phi-\varphi_0)+(U_l-U_r) N^2 t/2\right]\right\}
\ ,
\label{husimi2}
\eeq
which vanishes periodically in time with periodicity $T$.

{\it Interferometry:} The periodic time dependence for $t>t_e$ can be used for interferometry.
If the particles in the isolated wells experience different conditions
(e.g., different parameters $U_l$ and $U_r$ or different potentials), the expectation value 
$\langle\Psi_t|A|\Psi_t\rangle$ in Eq. (\ref{expect_2}) or the Husimi--$Q$ function 
in (\ref{husimi2}) would indicate this difference in the form of time-dependent oscillations.
In the case of different interaction parameters $U_l$ and $U_r$, for example, these oscillations appear
with periodicity $T=4\pi/(U_l-U_r) N^2$ due to interference effects.
Thus, the two wells can be considered as the two arms of a Mach-Zehnder interferometer. 
Dephasing strongly affects the performance, though, since a loss of particles
immediately destroys the PNS. For instance, if a particle is removed from one well
we obtain no interference, since
$
A a_r |\Psi_t\rangle = 0
$.
Therefore,
this interferometer would only work on time scales shorter than the dephasing time $t_d$.
On the other hand, the time scale $T=4\pi/(U_l-U_r) N^2$, required to measure the interference, 
decreases with $N^{-2}$ in Eq. (\ref{expect_2}) and Eq. (\ref{husimi2}). This would allow us to reduce
the real time of the measurement for obtaining a reasonable resolution when we increase the number 
of particles $N$.  
In other words, the interference at a given time interval $\Delta t$ measures $(U_l-U_r) N^2$ rather
than $U_l-U_r$. Thus, the precision for interferometric measurements of the nonlinear properties
is increased by a factor of $N^2$.

 
\begin{figure}
\psfrag{c_0,c_N}{$|c_0|$, $|c_N|$}
\psfrag{'00.dat' using 1:2}{$|c_0|$}
\psfrag{'00.dat' using 1:3}{$|c_N|$}
\psfrag{'75.dat' using 1:2}{$|c_0|$}
\psfrag{'75.dat' using 1:3}{$|c_N|$}
\begin{center}
\includegraphics[width=7.5cm,height=7cm]{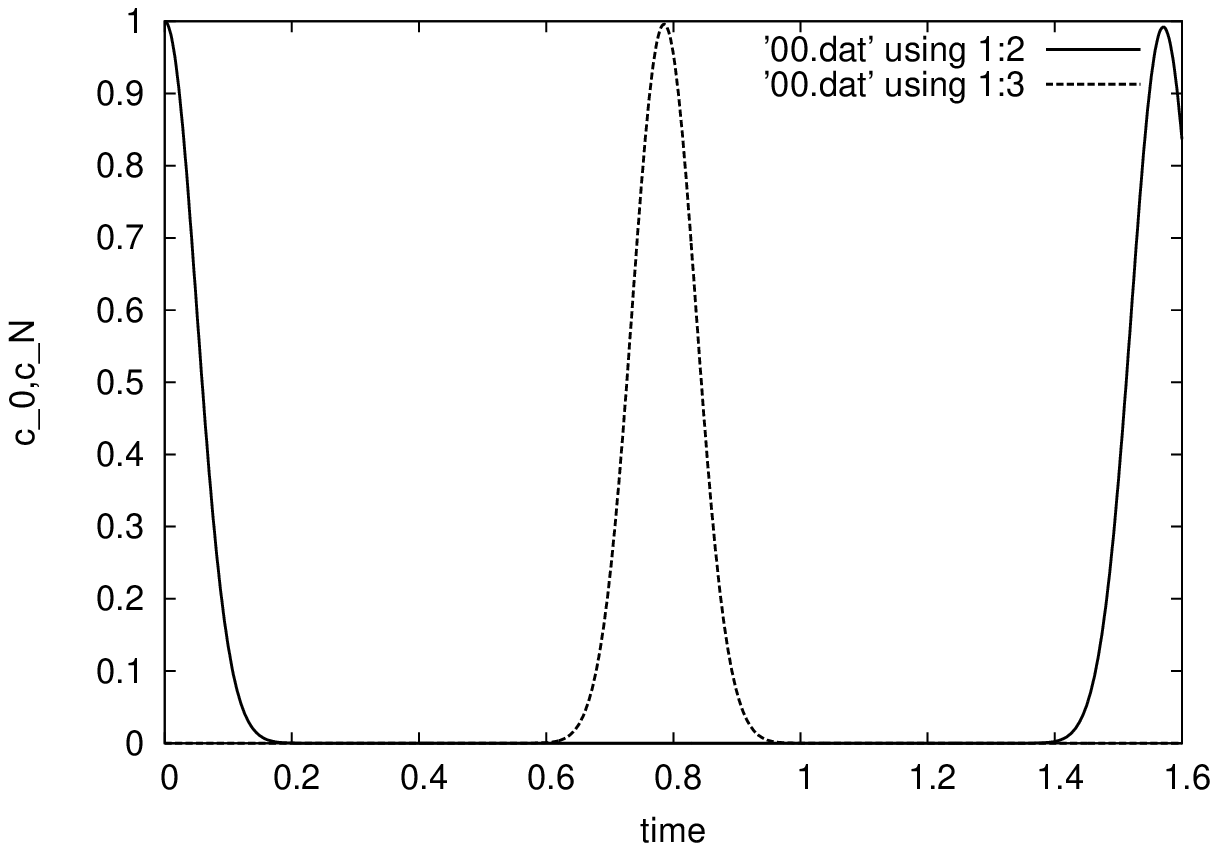}
\includegraphics[width=7.5cm,height=7cm]{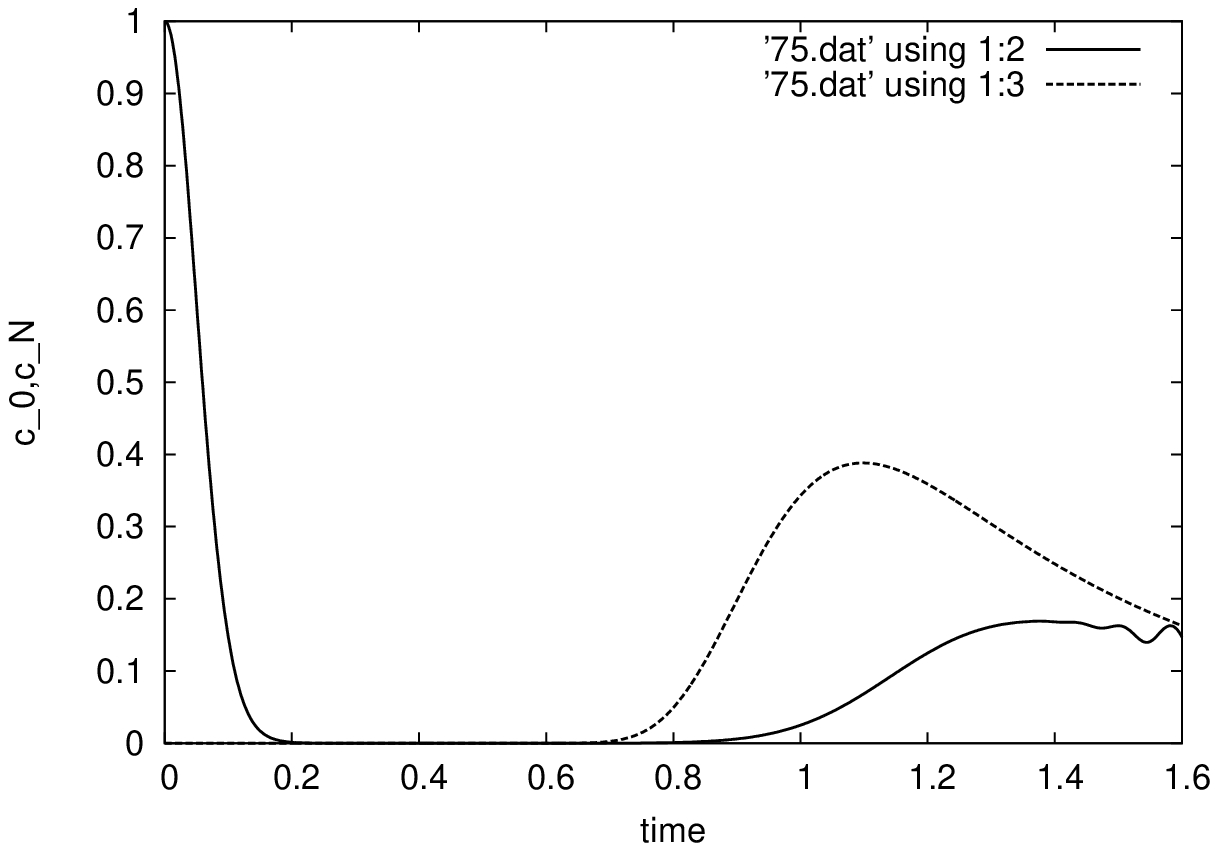}
\caption{
Dynamics of $|c_0|$ and $|c_N|$ with tunneling rate $J=4$ Hz
for $N=100$ bosons with $u\equiv NU/J=0$ (left) and $u=3.75$ (right).
There is a periodicity $T=\pi/2$ sec ($\approx 1.57$ sec) in the left plot. The steepness of the initial decay, 
which is the same for both systems, is determined by the number of bosons: $|c_0|\sim \exp(-NJ^2t^2/2)$.
}
\label{fig:dynamics1}
\end{center}
\end{figure}

\begin{figure}
\psfrag{c_0,c_N}{$|c_0|$, $|c_N|$}
\psfrag{'08.dat' using 1:2}{$|c_0|$}
\psfrag{'08.dat' using 1:3}{$|c_N|$}
\psfrag{'85.dat' using 1:2}{$|c_0|$}
\psfrag{'85.dat' using 1:3}{$|c_N|$}
\begin{center}
\includegraphics[width=7.5cm,height=7cm]{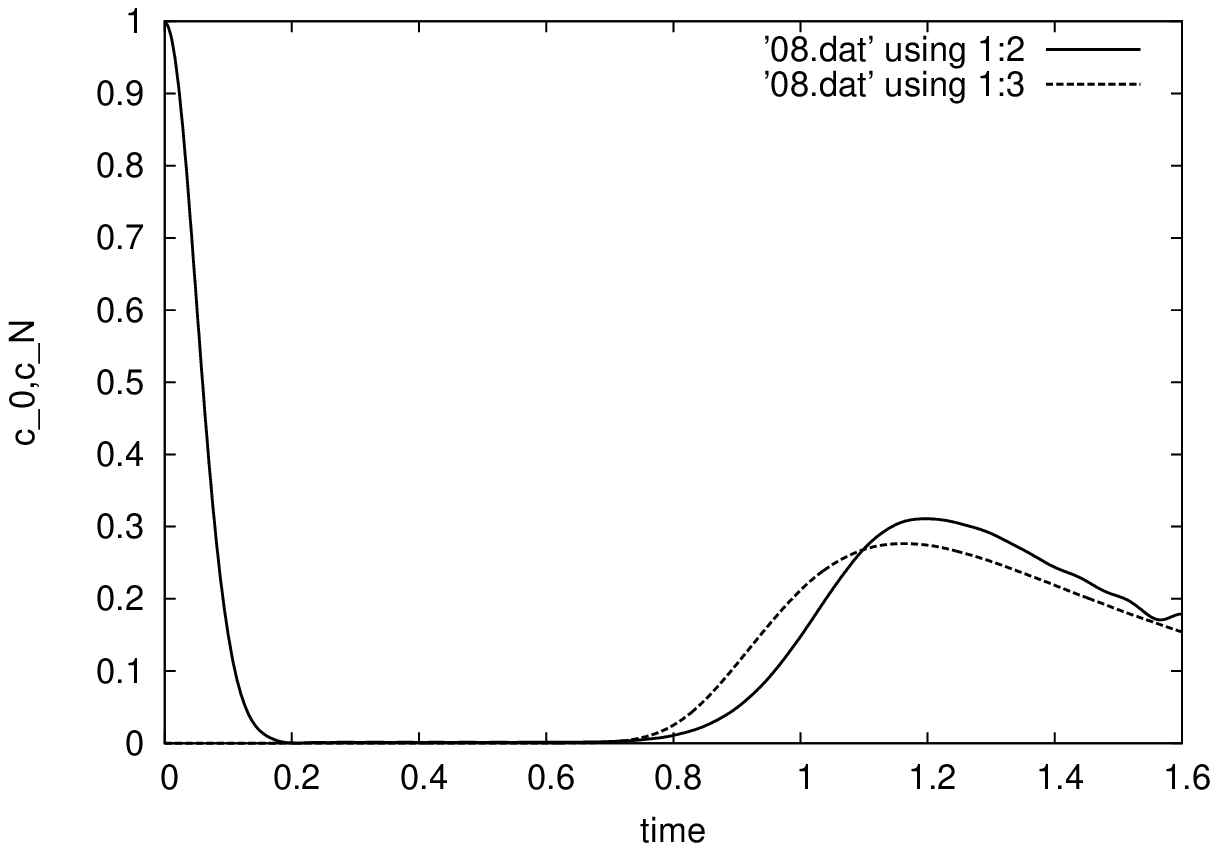}
\includegraphics[width=7.5cm,height=7cm]{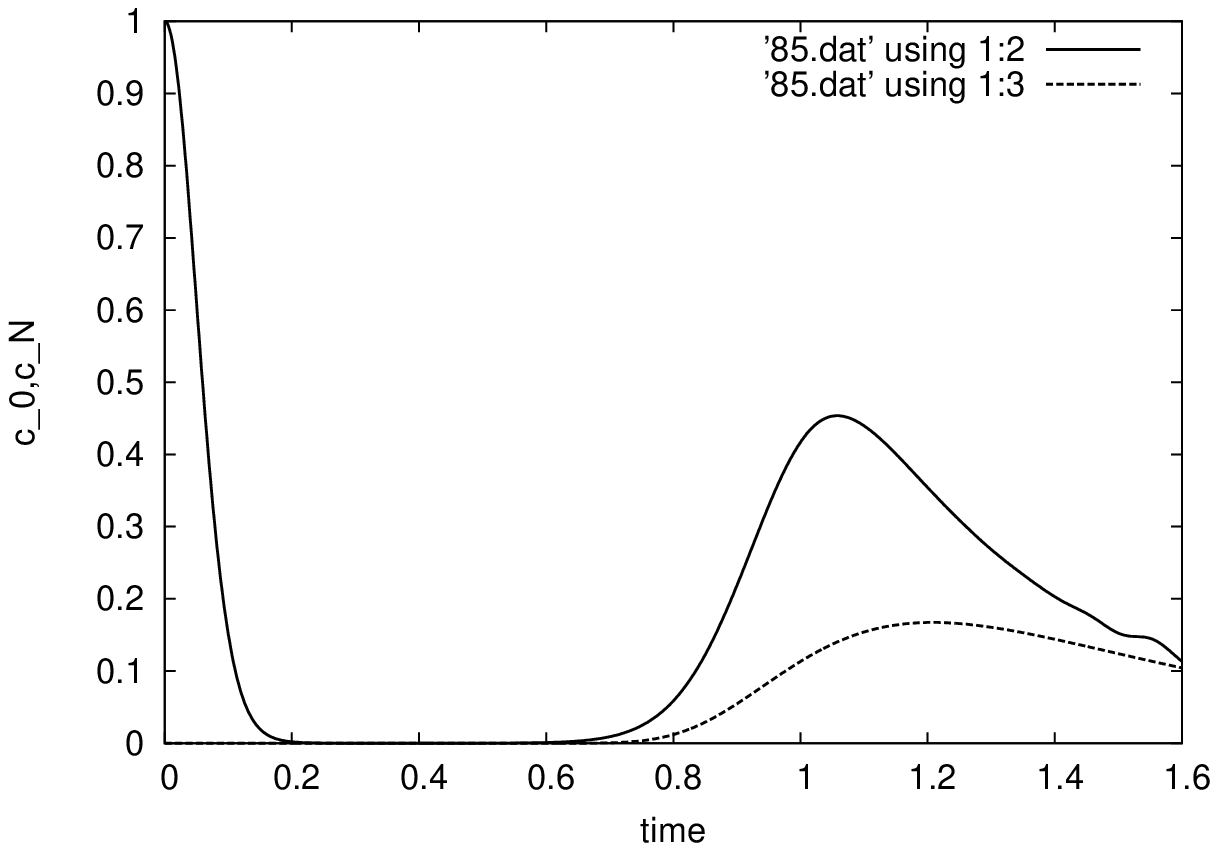}
\caption{
Dynamics of $|c_N|$ and $|c_0|$ 
for $N=100$ bosons with $u=4$ (left) and $u=4.25$ (right).
}
\label{fig:dynamics2}
\end{center}
\end{figure}

\begin{figure}
\psfrag{'60.dat' using 1:4}{$N=120$}
\psfrag{'50.dat' using 1:4}{$N=100$}
\psfrag{'40.dat' using 1:4}{$N=80$}
\psfrag{'30.dat' using 1:4}{$N=60$}
\psfrag{'20.dat' using 1:4}{$N=40$}
\psfrag{'10.dat' using 1:4}{$N=20$}
\psfrag{entanglement}{$p_e$}
\psfrag{max. entanglement}{$P_e$}
\begin{center}
\includegraphics[width=7.5cm,height=7cm]{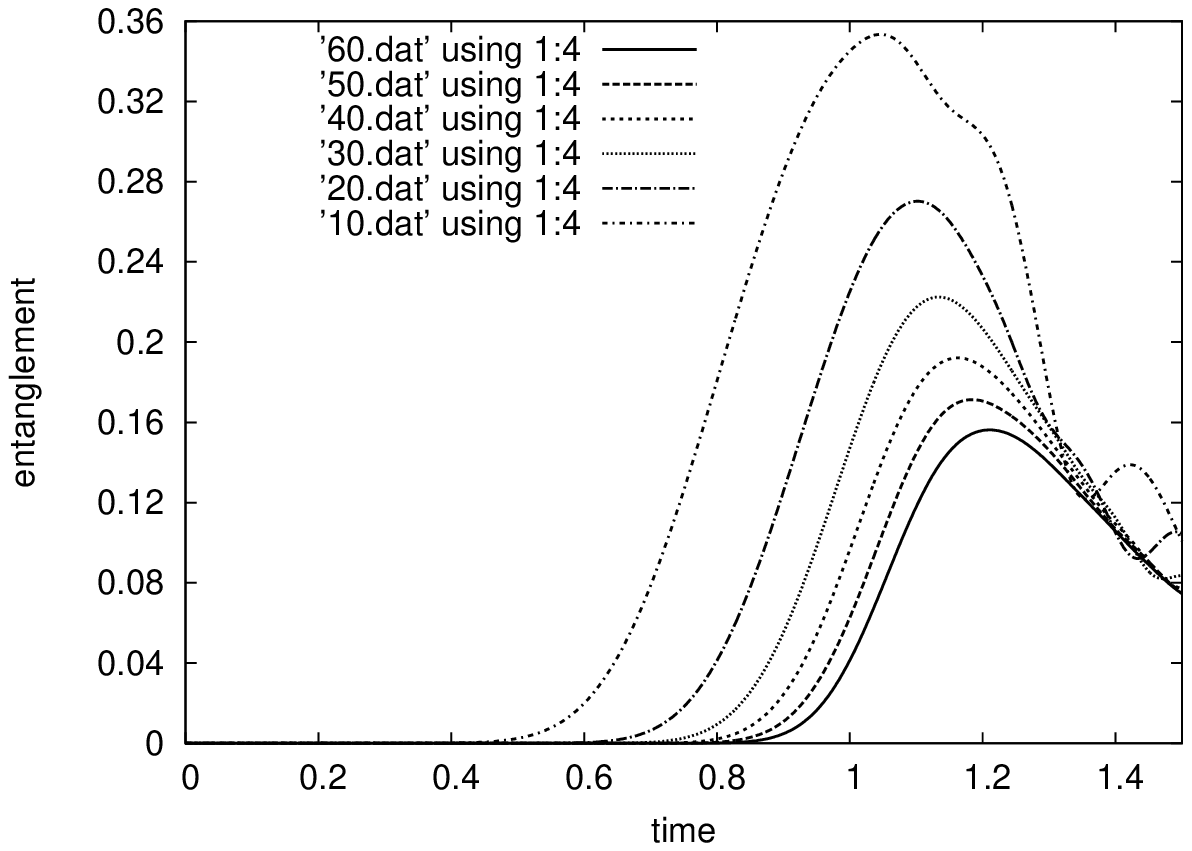}
\includegraphics[width=7.5cm,height=7cm]{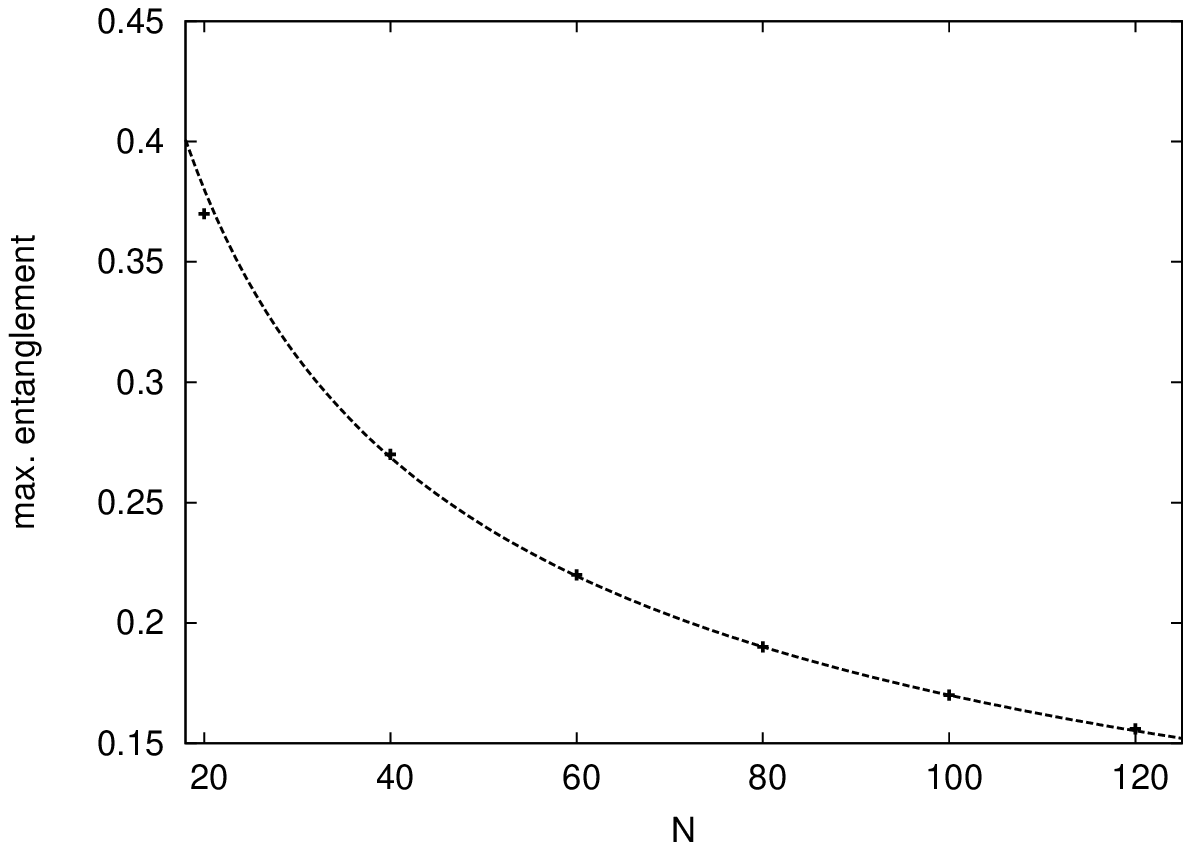}
\caption{
Left: Dynamics of entanglement $p_e(t)$ for different $N$ (left) with $u=4$. 
Right: Maximal entanglement $P_e$ as a function $N$ with the fitting curve $1.7 N^{-1/2}$.
} 
\label{fig:dynamics3}
\end{center}
\end{figure}

\begin{figure}
\psfrag{'425_100.dat' using 1:5}{$u=4.25$}
\psfrag{'40_100.dat' using 1:5}{$u=4.0$}
\psfrag{'375_100.dat' using 1:5}{$u=3.75$}
\psfrag{'425_100.dat' using 1:6}{$u=4.25$}
\psfrag{'40_100.dat' using 1:6}{$u=4.0$}
\psfrag{'375_100.dat' using 1:6}{$u=3.75$}
\psfrag{entanglement}{$\langle\Psi_t|A|\Psi_t\rangle$}
\begin{center}
\includegraphics[width=7.5cm,height=7cm]{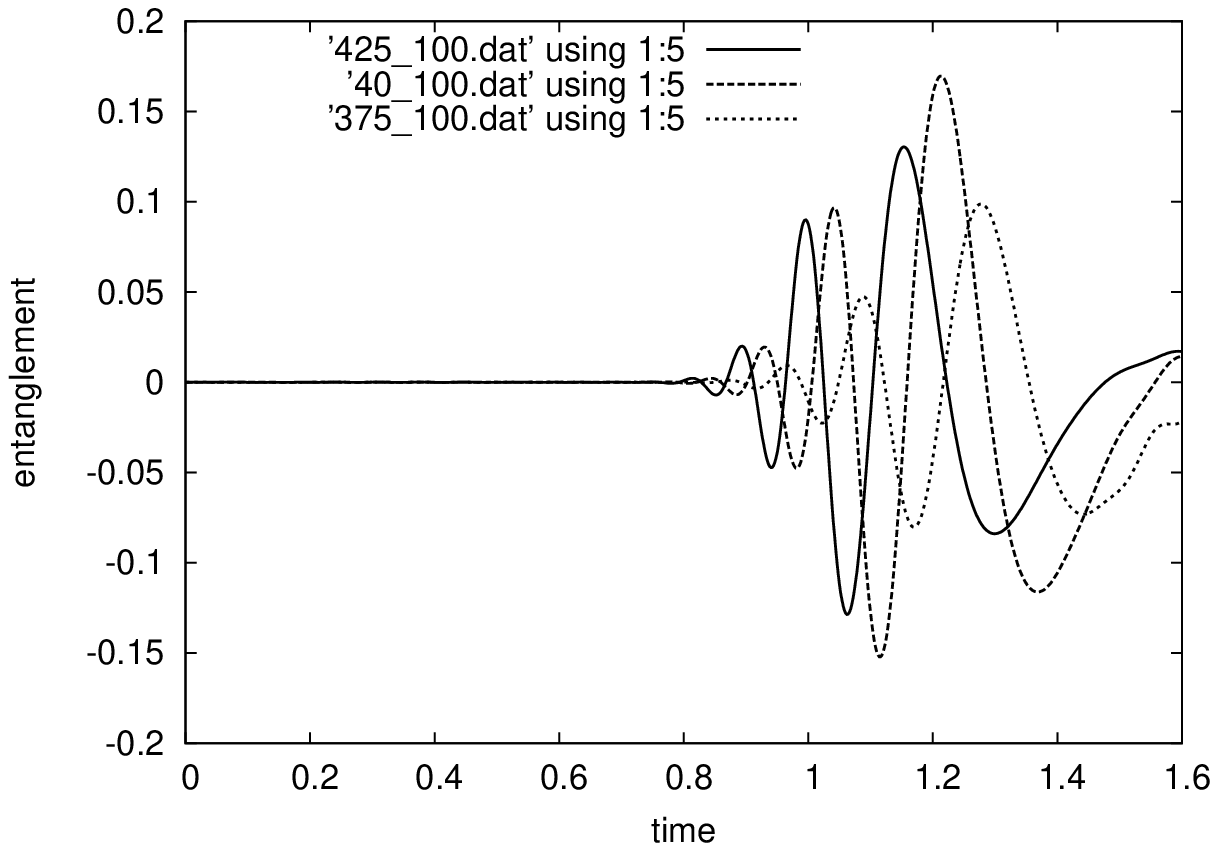}
\includegraphics[width=7.5cm,height=7cm]{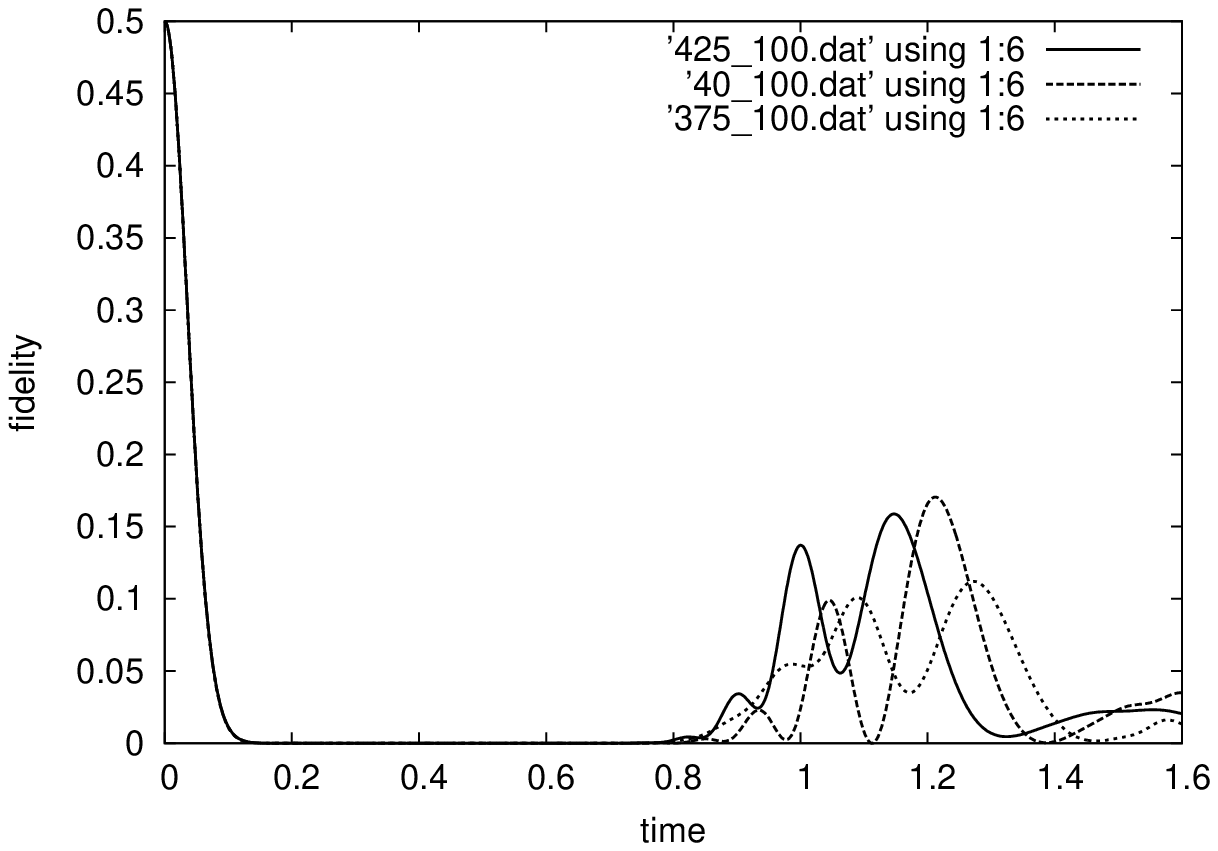}
\caption{
Dynamics of the coherence term $\langle\Psi_t|A|\Psi_t\rangle$ (left) and the fidelity $|\langle N00N|\Psi_t\rangle|^2$
(right) for $N=100$ with $J=4$ Hz and with different values of $U$. 
} 
\label{fig:dynamics4}
\end{center}
\end{figure}

\section{Conclusion}

The bosonic Josephson tunneling junction can be used as a device to create entangled states from a pure Fock state
by an evolution with the Hamiltonian (\ref{ham00}): 
\beq
|0,N\rangle \to \sum_{k=0}^N c_k e^{-iU_l k^2 t/2}|k\rangle\otimes e^{-iU_r (N-k)^2 t/2}|N-k\rangle
\ ,
\eeq
where the pure Fock state was prepared at time $t=0$ and the entangled PNS was created at an appropriate time $t_e>0$.
The new state on the right-hand side evolves for $t>t_e$,
which gives a constant value for $p_e$ and periodic functions in time for the expectation value
$\langle\Psi_t|A|\Psi_t\rangle$ of (\ref{expect_2}) and the Husimi--$Q$ function of (\ref{husimi2}).

For non-interacting bosons (e.g., for two coupled photonic cavities) the maximal entanglement $P_e$
decays exponentially with the particle number $N$, whereas for optimal interaction 
strength $u$ and optimal time $t_e$ the decay is $N^{-1/2}$, as depicted in Fig. \ref{fig:dynamics3}.
$t_e$ is in the time interval $[\pi/J,2\pi/J]$, independent of the interaction, whereas the optimal 
entanglement appears for the interaction strength $u\approx 4$. In this regime the collapse and revival
behavior has not fully developed and the moderate interaction favors the simultaneous occupation of
$|0,N\rangle$ and $|N,0\rangle$. Another important point is that the initial state $|0,N\rangle$ has
a high energy for the Hamiltonian $H$. Since the evolution re-distributes the energy to different
states, it is possible that the other high-energy state $|N,0\rangle$ will be occupied too. In contrast,
if the initial state is a low-energy state (e.g., $|N/2,N/2\rangle$), the formation of the N00N state 
would be much less likely.

After switching off the tunneling, the state can evolve without changing the entanglement, as long as no intervention
from outside or a loss of particles happen. Since the evolution is described by a phase inside each well, 
the two wells can be used as an interferometer with effective periodicity $T=4\pi/(U_l-U_r) N^2$ to measure
the difference of the particle interaction $U_r$, $U_l$ in the two wells.

\end{document}